\documentclass[aps,prb,twocolumn,showpacs,superscriptaddress]{revtex4}
\usepackage{amssymb,amsmath}
\usepackage{graphicx}
\usepackage{multirow}
\usepackage{hyperref}
\usepackage{appendix}
\usepackage{mathtools}
\usepackage{amsmath}
\usepackage{xcolor}
\usepackage{siunitx}
\graphicspath{{./IMG/}}

\begin{document}

\title{
Variational Polaron Equations Applied to the Anisotropic Fr{\"o}hlich Model}

\author{Vasilii Vasilchenko}
\affiliation{Skolkovo Institute of
Science and Technology, Skolkovo Innovation Center, Nobel St. 3, Moscow, 143026, Russia.}
\author{Andriy Zhugayevych}
\affiliation{Skolkovo Institute of
Science and Technology, Skolkovo Innovation Center, Nobel St. 3, Moscow, 143026, Russia.}
\author{Xavier Gonze}
\affiliation{Skolkovo Institute of
Science and Technology, Skolkovo Innovation Center, Nobel St. 3, Moscow, 143026, Russia.}
\affiliation{European Theoretical Spectroscopy Facility, Institute of Condensed Matter and Nanosciences, Universit\'{e} catholique de Louvain, Chemin des \'{e}toiles 8, bte L07.03.01, B-1348 Louvain-la-Neuve, Belgium}


\begin{abstract}
Starting from recent advances in the first-principles modeling of polarons, variational polaron equations in the strong-coupling adiabatic approximation are formulated in Bloch space.
In this framework, polaron formation energy as well as individual electron, phonon and electron-phonon contributions are obtained.
We suggest an efficient gradient-based optimization algorithm and apply these equations to the generalized Fr{\"o}hlich model with anisotropic non-degenerate electronic bands, both in two- and three-dimensional cases.
The effect of the divergence of Fr{\"o}hlich electron-phonon matrix elements at $\Gamma$-point is treated analytically, improving the convergence with respect to the sampling in reciprocal space.
The whole methodology is validated by obtaining the known asymptotic solution of the standard Fr{\"o}hlich model in isotropic scenario and also by comparing our results with the Gaussian ansatz approach, showing the difference between the numerically exact and Gaussian trial wavefunctions. Additionally, decomposition of the energy into individual terms allows one to recover the Pekar's 1:2:3:4 theorem, which is shown to be valid even in the anisotropic case.
We expect that the improvements in the formalism and numerical implementation will be applicable beyond the large polaron hypothesis inherent to Fr{\"o}hlich model.
\end{abstract}


\maketitle

\section{Introduction}
\label{sec:intro}

A polaron is a quasiparticle formed when an electron or a hole couple to the lattice vibrations of a system.
Polaron formation leads to effective mass renormalization, lattice deformation and potential self-trapping of the carrier.
This process occurs in various classes of materials: organic crystals \cite{Giannini2019}, perovskites \cite{Park2018, Ghosh2020}, oxides \cite{Brunin2019}, metal-ion storage materials \cite{Dathar2011}, 2D semiconductors \cite{Kang2018, Vasilchenko2021}.
Since many of these systems find their applications in electronics, and polarons in turn affect their optoelectronic properties, accurate prediction of polaronic effects is important.
Indeed, many recent state-of-the-art experimental and theoretical studies are devoted to the polaron physics \cite{Franchini2021} and encourage further investigations in this field. 

A polaron has several characteristics, like its formation energy, effective mass, mobility and optical response.
The first approach for the determination of polaron properties was suggested by Landau and Pekar \cite{Landau1948}. 
They made the hypothesis of large spatial extension of the quasiparticle compared to the lattice periodicity (large polaron), hence neglecting  the atomic details of the crystal.
They also treated the lattice deformation using classical mechanics, the deformation adjusting instantaneously and self-consistently to the charge carrier density. We will refer to this second hypothesis as the strong-coupling adiabatic approximation.
In this approach a charge carrier can be trapped in the deformation field it induces, a phenomenon called ``self-trapping".

Two other noticeable approaches were introduced later by Fr{\"o}hlich \cite{Frhlich1950} and Holstein \cite{Holstein1959a, Holstein1959b}. 
The Fr{\"o}hlich model takes into account coupling of a single electron to a dispersionless longitudinal optical (LO) phonon mode and also washes out the atomic details with a continuum approximation, describing large polarons \cite{Sendner2016, Zheng2019}, like Landau and Pekar approach, albeit with a quantum treatment of such phonon modes.
The Fr{\"o}hlich model has been the subject of sustained attention for several decades. 
One can treat rather easily two regimes, depending on the strength of the electron-phonon coupling: the above-mentioned strong-coupling limit, yielding a self-trapped state, and the weak-coupling limit, that can be treated using perturbation theory \cite{Froehlich1954}. Fr{\"o}hlich polarons in the weak-coupling scenario can also be described with variational formalism based on canonical transformation as in Lee-Low-Pines theory \cite{Lee1953}.
The treatment of the intermediate regime is more challenging, and can be done approximately using e.g. Feynman's path integral approach \cite{Feynman1955}, or by diagrammatic Monte Carlo techniques \cite{Mishchenko2000}.
The Holstein approach is a lattice model that considers local couplings to intramolecular vibrations describing small polarons and is commonly used to study polaronic properties of organic semiconductors \cite{Perroni2011, Zhugayevych2015}.

These models however are not able to describe the full complexity of real materials, with arbitrary degenerate electronic and phonon dispersion and complex electron-phonon interactions.  
Such features in general can only be taken into account by first-principles calculations, like Density Functional Theory (DFT). 
 DFT-based calculations of polaron properties have already shown their predictive power for describing polarons either in weak-coupling \cite{Marini2008, Giustino2010, Gonze2011a, Antonius2014, Ponce2014, Verdi2015, Ponce2015, Antonius2015, Giustino2017, Nery2018, Miglio2020, Neaton2020} or strong-coupling regimes \cite{Sio2019, Stoneham2007, Varley2012, Sadigh2015, Geneste2017, Kokott2018, Tantardini2022a}.
In the weak-coupling regime, the polaron formation energy is often termed
the zero-point renormalization of the electronic band edge energy.
 The electronic structure, i.e. Kohn-Sham (KS) states and eigenenergies in the DFT case, is affected by the atomic displacements, either perturbatively, in the weak-coupling case, or self-consistently, in the strong-coupling case. 
However, first-principles calculations in the strong-coupling regime are rather limited in the number of atoms that can be treated. 
Besides, intermediate coupling strength region is more challenging to address with first-principles approaches, while a large number of interesting materials might resort to such regime.

A step towards unified modeling of polarons in Bloch space, fully based on first principles, has recently been made by Sio, Verdi, Ponc\'e and Giustino (SVPG) \cite{Sio2019a, Sio2019}. 
Within some well-justified approximations, they derive from first principles a model Hamiltonian, in real space, then transform it into the basis of KS states and phonon modes. 
The SVPG approach is more general than both Fr{\"o}hlich and Holstein ones in the sense that it is applicable to both small and large polarons. 
It includes electronic, phonon and electron-phonon terms,
albeit for a single phonon-dressed charge carrier, like in these models. 
After having reduced the many-electron problem to a one-electron problem, Sio and coworkers solve this Hamiltonian in the strong-coupling adiabatic approximation.
Another formalism applicable to first-principles Hamiltonians is based on canonical transformation allowing for efficient consideration of nonadiabatic effects \cite{Hannewald2004, Lee2021, Luo2022}. 
Both aforementioned approaches can provide a foundation for further developments in the field of polaron physics.

In the present work, we reexamine the SVPG formalism, provide several improvements, and apply it to the generalized Fr{\"o}hlich model, to compare with known reference data.
The generalized Fr{\"o}hlich model introduced by Miglio \textit{et al.} in Ref.~\citenum{Miglio2020} and later examined by Guster \textit{et al.} in Ref.~\citenum{Guster2021} extends the original Fr{\"o}hlich model by taking into account degeneracy and anisotropy of electronic bands and their coupling to several LO phonon branches. 
It can be considered in both weak and strong coupling limits,
nevertheless retaining the large polaron hypothesis (ignoring atomic details).
Such generalization captures the essential effects of electron-phonon interaction in oxides and II-VI materials as shown by Miglio \textit{et al.} in Ref.~\citenum{Miglio2020}. 
Guster and coworkers further investigated this model using a Gaussian ansatz approach for a set of cubic materials, determining effective masses and localization lengths of polarons in the weak-coupling and strong-coupling limits respectively. 
However, the trial Gaussian wavefunction used in the strong-coupling limit may not be sufficient for a polaron wavefunction approximation when electronic bands are highly anisotropic, and in this case self-consistent methods may yield more accurate results.

The present work utilizes the aforementioned SVPG theoretical framework~\cite{Sio2019a, Sio2019}. By reformulating their approach we first derive general variational equations for polaronic energy, in the basis of KS states and vibrational eigenmodes, and suggest efficient minimization algorithms. 
Then we apply our approach to the generalized 2D and 3D Fr{\"o}hlich model in the adiabatic strong-coupling limit considering only the anisotropic non-degenerate electronic bands as extension to the standard Fr{\"o}hlich model. 
We also provide special treatment for the Fr{\"o}hlich electron-phonon matrix elements at $\Gamma$-point similar to Miglio \textit{et al.}, Ref.~\citenum{Miglio2020}, as in common representation they diverge at band edges.
The results are compared with the known asymptotic solution of the classic (isotropic) Fr{\"o}hlich model in the strong-coupling limit as well as with the adiabatic Gaussian ansatz approach suggested in Ref.~\citenum{Guster2021}. 
In particular, we find that the Gaussian ansatz delivers an accurate polaron formation energy for the whole range of anisotropy parameter, be it in 2D or 3D. 
In passing, we perform the decomposition of the energy of the Fr{\"o}hlich model in the strong-coupling limit into
individual terms and recover the Pekar’s 1:2:3:4 theorem, which we show to be valid even in the
anisotropic case. 

The paper is structured as follows.
In the next section, we give some background material and also provide notations for the reminder of the paper. Explicitly, we recall the SVPG approach (with an additional discussion of the choice of phase), and give an account of the standard (isotropic) Fr{\"o}hlich model with several well-established results.
Then, in Sec.~\ref{methods}, we present our methodological advances, namely (i) a generalization of SVPG approach which yields variational polaron equations in Bloch space, and is free of the phase convention on which SVPG relied, (ii) the formulation of the anisotropic Fr{\"o}hlich model with these variational polaron equations,  (iii) a treatment of the infrared divergence of the electron-phonon coupling, and (iv) the formulation of a preconditioned conjugate gradient algorithm to address the variational polaron equations in Bloch space.
In Sec.~\ref{result}, we deal with the 2D- and 3D-Fr{\"o}hlich model, isotropic and anisotropic using the above-mentioned algorithm. 
We analyze numerical aspects of this treatment as well as the physical characteristics of the polaron.  
Three appendices focus on the choice of phase for electron-phonon equations, on Pekar's 1:2:3:4 theorem, and on the scaling of the minimization algorithm.

\section{Background}
\label{sec:theory}

\subsection{First-principles modeling of a polaron}

The SVPG approach~\cite{Sio2019,Sio2019a} starts from the DFT total energy of a semiconductor or an insulator with the fully occupied valence bands and empty conduction bands.
Upon addition or removal of a single electron from a system, the charge change is in one well-defined spin channel, the system is thus spin-polarized, but SVPG assumes that the change in the overall electronic density is negligible.
The ground state total energy is expanded to the second order in displacements as 
\begin{equation}\label{eq:2}
\begin{aligned}
    & E[\{\psi_{n\mathbf{k}}\},\{\boldsymbol{\tau}_{\kappa p}\}] = E[\{\psi^0_{n\mathbf{k}}\},\{\boldsymbol{\tau}^0_{\kappa p}\}] \\
    & + \frac{1}{2}\sum_{\substack{\kappa\alpha p \\ \kappa' \alpha' p'}} C^0_{\kappa \alpha p,\kappa' \alpha' p'} \Delta \tau_{\kappa \alpha p} \Delta \tau_{\kappa' \alpha' p'} + \mathcal{O}(\Delta \tau^3),
\end{aligned}
\end{equation}
where $\psi_{n\mathbf{k}}$ are the KS wavefunctions of the occupied states, and general ionic coordinate of an atom $\kappa$ in a $p$-th supercell along the $\alpha$-direction  ${\tau}_{\kappa \alpha p}={\tau}_{\kappa \alpha p}^0+\Delta{\tau}_{\kappa \alpha p}$ is written as displacement $\Delta {\tau}_{\kappa \alpha p}$ from the equilibrium atomic position ${\tau}_{\kappa \alpha p}^0$, and $C^0_{\kappa \alpha p,\kappa' \alpha' p'}$ is the matrix of interatomic force constants \cite{Gonze1997, Giustino2017}.
The wavefunctions at distorted geometries $\psi_{n\mathbf{k}}$ are self-consistently optimized to minimize the total energy, so they are functions of $\boldsymbol{\tau}_{\kappa p}$, although this dependence is not mentioned explicitly in Eq.~\ref{eq:2}.
The energy is defined for a Born-von Karman (BvK) supercell containing $N_p$ unit cells.
The derivation of the formalism for electron and hole polarons is symmetrical and we will follow the authors' approach describing further only the electron polaron case.

Eq. (\ref{eq:2}) is then combined with the expression for the DFT total energy to obtain the DFT functional of a polaron with the associated wavefunction $\psi$, for which the change of electronic density is $\psi^*\psi = \Delta \rho$, which is negligible by the author's assumption ($\Delta \rho \ll \rho$):
\begin{equation}\label{eq:3}
\begin{aligned}
    E_\text{p}[\psi, \{\Delta {\tau}_{\kappa \alpha p}\}] & = E[\{\psi^0_{n\mathbf{k}}\},\{\boldsymbol{\tau}^0_{\kappa p}\}] \\
    & + \dfrac{1}{2} \sum_{\substack{\kappa\alpha p \\ \kappa' \alpha' p'}} C^0_{\kappa \alpha p,\kappa' \alpha' p'} \Delta \tau_{\kappa \alpha p} \Delta \tau_{\kappa' \alpha' p'} \\
    & + \int d\mathbf{r}\psi^*(\mathbf{r}) \hat{H}_\text{KS}[\rho, \{\boldsymbol{\tau}_{\kappa p}\}] (\mathbf{r}) \psi(\mathbf{r}).
\end{aligned}
\end{equation}
The third term describes electron and electron-phonon parts of the energy and contains the KS Hamiltonian of the system without addition of an electron, expanded up to the first order in $\Delta \tau_{\kappa\alpha p}$:
\begin{equation}\label{eq:4}
\begin{aligned}
     & \hat{H}_\text{KS}[\rho, \{\boldsymbol{\tau}_{\kappa p}\}] = \hat{H}_\text{KS}[\rho^0, \{\boldsymbol{\tau}^0_{\kappa p}\}] \\
     & + \sum_{\kappa\alpha p} \dfrac{\partial V^0_\text{KS}}{\partial \tau_{\kappa \alpha p}} \Delta \tau_{\kappa \alpha p},
\end{aligned}
\end{equation}
where $V^0_\mathrm{KS}$ denotes the KS self-consistent potential at equilibrium. 

The formation energy of the polaron is then obtained from the minimization of Eq.~(\ref{eq:3}) as 
\begin{equation}\label{eq:min}
\begin{aligned}
    \Delta E_\text{p} & = \mathrm{min}E_\text{p}[\psi, \{\Delta {\tau}_{\kappa \alpha p}\}]  \\ 
    & - \mathrm{min}E_\text{p}[\psi, \{\Delta {\tau}_{\kappa \alpha p}=0\}].
\end{aligned}
\end{equation}
The SVPG minimization formalism in real space is not presented here and can be found in the original paper. 
This yields self-consistent equations for the electron wavefunction and the atomic displacements, with a Lagrange multiplier associated to the norm conservation of the electronic wavefunction. 
Then, these equations are transformed to the basis of the Bloch electronic wavefunctions (KS basis in the DFT case) and phonon normal modes.

The polaronic wavefunction in the basis of the Bloch (KS) electronic wavefunctions reads as
\begin{equation}\label{eq:8}
    \psi(\mathbf{r}) = \frac{1}{\sqrt{N_p}}\sum_{n\mathbf{k}}A_{n\mathbf{k}}\psi^0_{n\mathbf{k}}(\mathbf{r})
\end{equation}
and must be normalized
\begin{equation}\label{eq:norm}
    \int d\mathbf{r}|\psi(\mathbf{r})|^2=1,
\end{equation}
so the electronic coefficients $A_{n\mathbf{k}}$ satisfy the following condition
\begin{equation}\label{eq:9}
    \dfrac{1}{N_p}\sum_{n\mathbf{k}}|A_{n\mathbf{k}}|^2=1.
\end{equation}
For an electron polaron, only unoccupied wavefunctions are used in Eq.~(\ref{eq:8}), while for a hole polaron, only occupied states are used. 

Atomic displacements in turn are expressed in terms of the phonon coefficients $B_{\mathbf{q}\nu}$ that represent the contribution of the normal modes to the displacements:
\begin{equation}\label{eq:10}
    \Delta \tau_{\kappa \alpha p} = - \dfrac{2}{N_p}\sum_{\mathbf{q}\nu}B^*_{\mathbf{q}\nu}\left(\dfrac{1}{2M_{\kappa}\omega_{\mathbf{q}\nu}}\right)^{1/2}e_{\kappa\alpha, \nu}(\mathbf{q})e^{i\mathbf{q}\cdot\mathbf{R}_p},
\end{equation}
where $M_\kappa$ denotes the mass of the $\kappa$ atom, $e_{\kappa\alpha, \nu}(\mathbf{q})$ is the orthonormal eigenmode of the corresponding phonon branch and $\mathbf{R}_p$ is a vector of the unit cell $p$ in real space.
$N_p$ is the number of primitive cells in a BvK supercell that hosts a polaron.
In Bloch space formulation this supercell is defined by the sampling of the Brillouin zone (BZ), namely, a uniform $\Gamma$-centred $\mathbf{k}$-point grid, with $N$ being the linear size of the grid.
In this sense for example a $10 \times 10 \times 10$ $\mathbf{k}$-point grid in reciprocal space corresponds to an equivalent $10 \times 10 \times 10$ supercell in real space.

Electronic $A_{n\mathbf{k}}$ and phonon $B_{\mathbf{q}\nu}$ parts of the polaron formation are coupled through the electron-phonon matrix elements \cite{Giustino2017} that represent the probability of scattering of an electron from the state $\psi^0_{n\mathbf{k}}$ into the state $\psi^0_{m \mathbf{k}+\mathbf{q}}$ through a phonon of the branch $\nu$ with momentum $\mathbf{q}$:
\begin{equation}\label{eq:epi}
\begin{aligned}
    g_{mn\nu}(\mathbf{k}, \mathbf{q}) = \sum_{\kappa \alpha p} \left( \dfrac{1}{2M_\kappa \omega_{\mathbf{q}\nu}} \right)^{1/2} e_{\kappa\alpha, \nu}(\mathbf{q})e^{i\mathbf{q}\cdot\mathbf{R}_p} \\
    \times \int d\mathbf{r} \psi^{0*}_{m\mathbf{k}+\mathbf{q}}(\mathbf{r}) \dfrac{\partial V^0_\mathrm{KS}}{\partial \tau_{\kappa \alpha p}} \psi^0_{n\mathbf{k}}(\mathbf{r}).
\end{aligned}
\end{equation}

At this point it is worth mentioning SVPG relies on the time-reversal symmetry of the electron-phonon matrix elements for the electron-phonon Hamiltonian to be hermitian:
\begin{equation}\label{eq:bh}
    g_{mn\nu}(-\mathbf{k}, -\mathbf{q})=g^*_{mn\nu}(\mathbf{k}, \mathbf{q}).
\end{equation}
This relation comes from the Born and Huang convention, Eq. (24.18) of Ref.~\onlinecite{Born54}, for the symmetry of eigenmodes:
\begin{equation}
    e_{\kappa\alpha, \nu}(-\mathbf{q}) = e^*_{\kappa\alpha, \nu}(\mathbf{q}).
\end{equation}
Alternatively, one can use Leibfried (p. 104 of Ref. \onlinecite{Leibfried55}) convention:
\begin{equation}
    e_{\kappa\alpha, \nu}(-\mathbf{q}) = -e^*_{\kappa\alpha, \nu}(\mathbf{q}).
\end{equation}
The choice of convention affects the form of some relations in Section \ref{sec:fro}, see  Ref.~\onlinecite{Guster2022}.
In later sections dealing with the Fr\"ohlich model, we will use Born and Huang convention.
Nonetheless, $g_{mn\nu}(\mathbf{k}, \mathbf{q})$ obtained from first-principles calculation have gauge arbitrariness due to the arbitrary phase factor of KS wavefunctions and phonon eigenmodes that enter Eq. (\ref{eq:epi}), so none of the conventions remain valid, generally speaking. 
While SVPG suggests a computational framework to get a unique gauge for all these quantities and work with Born and Huang convention, in Section \ref{sec:var} we use an alternative approach to explicitly make the Hamiltonian hermitian. 
Additionally, general discussion on how arbitrary phase factor of $e_{\kappa\alpha, \nu}$ affects electron-phonon equations is provided in Appendix \ref{asec:phase}.

From Eqs. (\ref{eq:min})-(\ref{eq:epi}) the following self-consistent eigenvalue
problem in Bloch space is obtained:
\begin{equation}\label{eq:6}
    \dfrac{2}{N_p} \sum_{\mathbf{q}m\nu}B_{\mathbf{q}\nu}g^*_{mn\nu}(\mathbf{k}, \mathbf{q})A_{m\mathbf{k}+\mathbf{q}}=(\varepsilon_{n\mathbf{k}}-\varepsilon) A_{n\mathbf{k}},
\end{equation}
\begin{equation}\label{eq:7}
    B_{\mathbf{q}\nu}=\dfrac{1}{N_p}\sum_{mn\mathbf{k}}A^*_{m\mathbf{k+q}}\dfrac{g_{mn\nu}(\mathbf{k}, \mathbf{q})}{\omega_{\mathbf{q}\nu}}A_{n\mathbf{k}}.
\end{equation}  
Parameters that enter these equations are eigenergies $\varepsilon_{n\mathbf{k}}$ of the relevant KS states, phonon frequencies $\omega_{\mathbf{q}\nu}$ and electron-phonon matrix elements $g_{mn\nu}(\mathbf{k}, \mathbf{q})$.
All these quantities can be obtained from the first-principles calculations \cite{Giustino2017} and their initialization allows one to start a self-consistent iterative procedure to solve Eqs.~(\ref{eq:6}) and (\ref{eq:7}) for electron and phonon parts of the polaron wavefunction  $\boldsymbol{A}$ and $\boldsymbol{B}$ (here we use italicized bold symbols to denote a set of coefficients, i.e. $\boldsymbol{A} \equiv \{ A_{n\mathbf{k}} \}$ ), and polaron eigenvalue~$\varepsilon$.

Resulting polaron eigenvalue $\varepsilon$ can be interpreted as energy of the localized state in the bandgap once the polaron is formed, with $\varepsilon_\mathrm{CBM}$ being the conduction band minimum. Electron polaron formation energy in terms of $\boldsymbol{A}$, $\boldsymbol{B}$ is given as
\begin{equation}\label{eq:11}
\begin{aligned}
    \Delta E_\text{p}(\boldsymbol{A}, \boldsymbol{B}) & =  \dfrac{1}{N_p}\sum_{n\mathbf{k}}|A_{n\mathbf{k}}|^2(\varepsilon_{n\mathbf{k}} - \varepsilon_\mathrm{CBM}) \\
    & - \frac{1}{N_p}\sum_{\mathbf{q}\nu} |B_{\mathbf{q}\nu}|^2\omega_{\mathbf{q}\nu}.
\end{aligned}
\end{equation}
This equation is not variational and relies on iterative solution of Eqs. (\ref{eq:6}) and (\ref{eq:7}) with convergence criteria being the absolute difference between $\Delta E_\text{p}$ at consequent steps becoming lower than a certain threshold.

In this paper we will derive a variational expression for $\Delta E_p(\boldsymbol{A}, \boldsymbol{B})$, allowing for an employment of various iterative minimization algorithms \cite{Pulay1980, Pulay1982, Gonze1996, Woods2019}, more efficient than the one suggested in the original paper.

\subsection{Fr{\"o}hlich model}\label{sec:fro}

The Fr{\"o}hlich model, either standard or generalized, starts from the following Hamiltonian:
\begin{equation}\label{eq:27}
    \hat{H}^\text{Fr} = \hat{H}^\text{Fr}_\text{e} + \hat{H}^\text{Fr}_\text{ph} + \hat{H}^\text{Fr}_\text{el-ph},
\end{equation} 
which, similar to Eq. (\ref{eq:3}), has electron, phonon and electron-phonon terms that contribute to the total energy of a system.

The original model suggested by Fr{\"o}hlich implies the following approximations: (i) there is one isotropic electronic band with effective mass $m^*$, (ii) coupling is considered only to one dispersionless LO phonon mode with frequency $\omega_\text{LO}$, (iii) the  character of a crystal is ignored and one deals with continuum limit. In this simplified scenario the terms of Eq. (\ref{eq:27}) become
\begin{equation}\label{eq:fr_el}
    \hat{H}^\text{Fr}_\text{e} = \sum_{\mathbf{k}}\dfrac{\mathbf{k}^2}{2m^*}\hat{c}^\dagger_{\mathbf{k}}\hat{c}_{\mathbf{k}},
\end{equation}
\begin{equation}
    \hat{H}^\text{Fr}_\text{ph} = \sum_{\mathbf{q}}\omega_\text{LO}\hat{a}^\dagger_{\mathbf{q}}\hat{a}_{\mathbf{q}},
\end{equation}
\begin{equation}\label{eq:fr_elph}
    \hat{H}^\text{Fr}_\text{el-ph} = \sum_{\mathbf{k}\mathbf{q}} g^\text{Fr}(\mathbf{q})\hat{c}^\dagger_{\mathbf{k}+\mathbf{q}}\hat{c}_{\mathbf{k}}(\hat{a}_{\mathbf{q}} + \hat{a}^\dagger_{\mathbf{-q}}),
\end{equation}
where $\hat{c}_{\mathbf{k}}$/$\hat{c}^\dagger_{\mathbf{k}}$ and $\hat{a}_{\mathbf{q}}$/$\hat{a}^\dagger_{\mathbf{q}}$  are the electron and phonon creation/annihilation operators respectively. Electron-phonon matrix elements are given as
\begin{equation}\label{eq:31}
g_\text{3D}^\text{Fr}(\mathbf{q}) = \dfrac{1}{|\mathbf{q}|}\left( \dfrac{2\pi \omega_\text{LO}}{N_p \Omega_0} {\epsilon^{*}}^{-1} \right)^{1/2}
\end{equation}
in the three-dimensional case, and as
\begin{equation}\label{eq:32}
g_\text{2D}^\text{Fr}(\mathbf{q}) = \dfrac{1}{|\mathbf{q}|^{1/2}}\left( \dfrac{\pi \omega_\text{LO}}{N_p \Omega_0} {\epsilon^{*}}^{-1} \right)^{1/2}
\end{equation}
in the two-dimensional case.\cite{Peeters1986}
This definition of the 2D Fr{\"o}hlich coupling has been the basis of several investigations, including e.g. a diagrammatic Monte Carlo reference study\cite{Hahn2018}. 
However, it corresponds to a idealized strictly 2D system. By contrast, the study of realistic systems with 2D characteristics embedded in 3D space, like free-standing monolayers or slabs, or even such systems deposited on surfaces, yield a different, much more complex, functional behaviour.\cite{Sohier2016, Sohier2017, Sohier2017a, Deng2021, Sio2022}. 
As our purpose is to compare our approach to reference data, we stick to the idealized 2D functional form Eq.~\ref{eq:32}.

In Eqs.~(\ref{eq:31}) and (\ref{eq:32}), we have followed the Born and Huang choice of phase, as explained in Appendix \ref{asec:phase}, and in Ref.~\onlinecite{Guster2022}.
Here $\epsilon^*$ is the effective permittivity due only to the ionic movements, defined from the difference of inverse high-frequency and static permittivities $\epsilon^{\infty}$ and $\epsilon^0$:
\begin{equation}
({\epsilon^{*}})^{-1} = ({\epsilon^{\infty}})^{-1} - ({\epsilon^{0}})^{-1}.
\end{equation}
Volume (area) of the 3D (2D) BvK supercell are denoted by $N_p \Omega_0$, where $\Omega_0$ is the volume (area) of the corresponding primitive cell.

It is convenient to introduce the dimensionless parameter, the so-called Fr{\"o}hlich coupling constant,
\begin{equation}
\alpha = \left( \dfrac{m^*}{2\omega_\text{LO}} \right)^{1/2}({\epsilon^{*}})^{-1}.
\end{equation}
Depending on the value of $\alpha$ the model has asymptotic solution. 
In the strong-coupling regime, one takes the limit $\alpha \rightarrow \infty$, and 
the polaron formation energy in 3D \cite{JMiyake1976} and 2D \cite{Xiaoguang1985} cases is expanded as
\begin{equation}\label{eq:35}
{\Delta E^\text{3D}_\text{p}} \approx -\omega_\text{LO}(0.1085\alpha^2+2.836+\mathcal{O}(1/\alpha^2)),
\end{equation}
\begin{equation}\label{eq:36}
{\Delta E^\text{2D}_\text{p}} \approx -\omega_\text{LO}(0.4047\alpha^2 + \mathcal{O}(\alpha^0)).
\end{equation}
The strong-coupling case is also captured by the variational approach \cite{Landau1948}, in which both electronic and phonon parts (displacements) of the polaron wavefunction are frozen self-consistently, and one works in the adiabatic approximation.

Alternatively, Fr{\"o}hlich model can be solved asymptotically in the weak-coupling regime ($\alpha \rightarrow 0$) with polaron formation energy showing leading linear dependence on~$\alpha$ \cite{Smondyrev1986, Fomin1994}.
Qualitative breakdown between the weak and strong coupling regimes occurs at $\alpha \approx 6$, and treatment of such intermediate coupling requires more sophisticated approaches \cite{Feynman1955, Mishchenko2000}.

Also, it has been shown \cite{Lemmens1973} that the 1:4 relation of the so-called 1:2:3:4 theorem of Pekar is valid for all ranges of~$\alpha$.
In adiabatic regime this theorem establishes the ratio between the effective kinetic energy of the electron trapped inside a polaron $E_\text{el}$, the lattice distortion energy $E_\text{ph}$, the energy of a localized polaronic state $\varepsilon$ and the electron-phonon interaction energy $E_\text{el-ph}$:
\begin{equation}\label{eq:37}
    E_\text{el} : E_\text{ph} : -\varepsilon : -E_\text{el-ph} = 1 : 2 : 3 : 4.
\end{equation}
Further discussion on this relation is also provided in Appendix~\ref{asec:pekar}.

In addition, some of the aforementioned restrictions can be bypassed by considering the generalized Fr{\"o}hlich model \cite{Miglio2020, Guster2021}. It allows to take into account degeneracy and anisotropy of the electronic bands and their coupling to several possible LO phonon modes instead of only one. The electron-phonon coupling then becomes more complex, but reduces to Eq. (\ref{eq:31}) and (\ref{eq:32}) not only in the limit of standard approximations, but also when electronic bands are parabolic and non-degenerate. 

\section{Theory and Implementation}
\label{methods}

\subsection{Variational polaron equations in Bloch space}\label{sec:var}

In order to formulate a variational expression for $\Delta E_\text{p}$ in Bloch space we start from Eq. (\ref{eq:3}), which is variational in real space under the normalization constraint Eq. (\ref{eq:norm}).

It is convenient to split the polaron energy into four parts, namely constant ground-state term and electron, phonon and electron-phonon terms:
\begin{equation}\label{eq:13}
    E_\text{p} = E^0 + E_\text{el} + E_\text{ph} + E_\text{el-ph},
\end{equation}
where
\begin{equation}\label{eq:14}
    E_\text{el} = \int d\mathbf{r}\psi^*(\mathbf{r}) \hat{H}_\text{KS}^0(\mathbf{r}) \psi(\mathbf{r}),
\end{equation}
\begin{equation}\label{eq:15}
    E_\text{ph} = \dfrac{1}{2} \sum_{\substack{\kappa\alpha p \\ \kappa' \alpha' p'}} C^0_{\kappa \alpha p,\kappa' \alpha' p'} \Delta \tau_{\kappa \alpha p} \Delta \tau_{\kappa' \alpha' p'},
\end{equation}
\begin{equation}\label{eq:16}
    E_\text{el-ph} = \int d\mathbf{r}\psi^*(\mathbf{r}) \sum_{\kappa\alpha p} \dfrac{\partial V^0_\text{KS} (\mathbf{r})}{\partial \tau_{\kappa \alpha p}} \Delta \tau_{\kappa \alpha p} \psi(\mathbf{r}).
\end{equation}
Such splitting will allow us to separately reformulate each part in terms of $\boldsymbol{A}$, $\boldsymbol{B}$ in Bloch space and later examine their individual contribution to the formation of a polaron.

Starting first with the electron part, we rely on the expansion of the polaron wavefunction $\psi$ in basis of KS states given by Eq. (\ref{eq:8}), keeping in mind that they are orthonormalized eigenfunctions of the KS Hamiltonian.
By combining Eqs. (\ref{eq:8}) and (\ref{eq:14}) one obtains
\begin{equation}\label{eq:17}
    E_\text{el} = \dfrac{1}{N_p} \sum_{n\mathbf{k}}|A_{n\mathbf{k}}|^2\varepsilon_{n\mathbf{k}},
\end{equation}
which is in agreement with Eq. (\ref{eq:11}).

For the phonon term we use the explicit expression for the matrix of interatomic force constants,
\begin{equation}\label{eq:18}
\begin{aligned}
    & C^0_{\kappa \alpha p,\kappa' \alpha' p'} = \\
    & \dfrac{\left(M_\kappa M_{\kappa'} \right)^{1/2}}{N_p}  \sum_{\mathbf{q} \nu}  e_{\kappa\alpha, \nu}(\mathbf{q}) \omega_{\mathbf{q}\nu}^2 e^*_{\kappa' \alpha', \nu}(\mathbf{q}) e^{i\mathbf{q}\cdot(\mathbf{R}_p - \mathbf{R}_{p'})},
\end{aligned}
\end{equation}
and Eq. (\ref{eq:10}) for the displacements $\Delta \tau_{\kappa \alpha p}$.
In order to be consistent with the Eq. (\ref{eq:11}) and obtain $|B_{\mathbf{q}\nu}|^2$ in the final result, we also note that $\Delta \tau_{\kappa \alpha p} = \Delta \tau^*_{\kappa \alpha p}$ since the displacements are real quantities.
Hence after combining Eqs. (\ref{eq:10}),  (\ref{eq:15}) and (\ref{eq:18}) the phonon part of the polaron energy reads as 
\begin{equation}\label{eq:19}
    E_\text{ph} = \frac{1}{N_p}\sum_{\mathbf{q}\nu} |B_{\mathbf{q}\nu}|^2\omega_{\mathbf{q}\nu}.
\end{equation}
This term comes into the variational equation with a different sign than in Eq. (\ref{eq:11}) and clearly indicates an increase of the polaronic energy due to the lattice deformation. 

To obtain the electron-phonon contribution to the energy, we substitute Eqs. (\ref{eq:8}) and (\ref{eq:10}) into Eq. (\ref{eq:16}) and use Eq. (\ref{eq:epi}). After some algebra the resulting term will be as follows
\begin{equation}\label{eq:21}
    E_\text{el-ph} = -\dfrac{2}{N_p}\sum_{nm\nu}\sum_{\mathbf{k}\mathbf{q}} A^*_{m\mathbf{k}+\mathbf{q}}A_{n\mathbf{k}}B^*_{\mathbf{q}\nu} g_{mn\nu}(\mathbf{k}, \mathbf{q}).
\end{equation}
At this point we recall the gauge arbitrariness of $g_{mn\nu}(\mathbf{k}, \mathbf{q})$ mentioned in the previous section. To tackle this problem we note that $E_\text{el-ph}$ has to be real, so it might me expressed alternatively by its complex conjugate or their average
\begin{equation}\label{eq:22}
\begin{aligned}
     & E_\text{el-ph} = \\
     & -\dfrac{1}{N_p}\sum_{nm\nu}\sum_{\mathbf{k}\mathbf{q}} \left( A^*_{m\mathbf{k}+\mathbf{q}}A_{n\mathbf{k}}B^*_{\mathbf{q}\nu} g_{mn\nu}(\mathbf{k}, \mathbf{q}) +
     (c.c.) \right).
\end{aligned}
\end{equation}
This expression has advantage to always be real regardless of electron and phonon parts of the polaron wavefunction and gauge arbitrariness of matrix elements.

Now from Eqs. (\ref{eq:13}), (\ref{eq:17}), (\ref{eq:19}) and (\ref{eq:22}) we get the sought variational expression for the total energy of polaron, also using Eq. (\ref{eq:9}) with the Lagrange multiplier term to impose the normalization condition on its wavefunction:
\begin{equation}\label{eq:23}
\begin{aligned}
    & E_\text{p} = E^0 + \dfrac{1}{N_p} \sum_{n\mathbf{k}}|A_{n\mathbf{k}}|^2\varepsilon_{n\mathbf{k}} \\
    & - \varepsilon \left( \dfrac{1}{N_p} \sum_{n\mathbf{k}}|A_{n\mathbf{k}}|^2 - 1 \right) + 
    \frac{1}{N_p}\sum_{\mathbf{q}\nu} |B_{\mathbf{q}\nu}|^2\omega_{\mathbf{q}\nu} \\
    & - \dfrac{1}{N_p}\sum_{nm\nu}\sum_{\mathbf{k}\mathbf{q}} \left( A^*_{m\mathbf{k}+\mathbf{q}}A_{n\mathbf{k}}B^*_{\mathbf{q}\nu} g_{mn\nu}(\mathbf{k}, \mathbf{q}) +
     (c.c.) \right).
\end{aligned}
\end{equation}
This is a central result of this paper.

This expression yields minimum conditions for the polaron formation energy $\Delta E_\text{p}$ by differentiation for $A_{n\mathbf{k}}$, $B_{\mathbf{q}\nu}$ under constraint Eq. (\ref{eq:9}), with obvious notations for their real or imaginary parts, respectively:
\begin{equation}\label{eq:24}
\begin{aligned}
    & 0 = \dfrac{\partial E_\text{p}(\boldsymbol{A}, \boldsymbol{B})}{\partial \operatorname{Re/Im}(A_{n'\mathbf{k}'})} = \dfrac{2}{N_p}\operatorname{Re/Im}(A_{n'\mathbf{k}'})(\varepsilon_{n'\mathbf{k}'} - \varepsilon)  \\
    & - \dfrac{2}{N_p^2}\sum_{n\nu\mathbf{q}} \operatorname{Re/Im}(A_{n\mathbf{k}'-\mathbf{q}}B^*_{\mathbf{q}\nu}g_{n'n\nu}(\mathbf{k}'-\mathbf{q}, \mathbf{q}) \\ 
    & + A_{n\mathbf{k}'+\mathbf{q}}B_{\mathbf{q}\nu}g^*_{nn'\nu}(\mathbf{k}', \mathbf{q})),
\end{aligned}
\end{equation}
\begin{equation}\label{eq:25}
\begin{aligned}
    0 = \dfrac{\partial E_\text{p}(\boldsymbol{A}, \boldsymbol{B})}{\partial \operatorname{Re/Im}(B_{\mathbf{q}'\nu'})} = \dfrac{2}{N_p}\operatorname{Re/Im}(B_{\mathbf{q}'\nu'})\omega_{\mathbf{q}'\nu'} \\
    - \dfrac{2}{N_p^2}\sum_{nm\mathbf{k}}\operatorname{Re/Im}(A^*_{m\mathbf{k}+\mathbf{q}'}A_{n\mathbf{k}}g_{mn\nu'}(\mathbf{k}, \mathbf{q}')).
\end{aligned}
\end{equation}
From Eqs. (\ref{eq:24}), (\ref{eq:25})  one obtains the eigenvalue problem similar to the one defined by Eqs. (\ref{eq:6}), (\ref{eq:7}). However, the result is more general, with Eq. (\ref{eq:6}) now becoming
\begin{equation}\label{eq:26}
\begin{aligned}
    \dfrac{1}{N_p} \sum_{\mathbf{q}m\nu}(& B_{\mathbf{q}\nu}g^*_{mn\nu}(\mathbf{k}, \mathbf{q})A_{m\mathbf{k}+\mathbf{q}} + \\
    & B^*_{\mathbf{q}\nu}g_{mn\nu}(\mathbf{k}-\mathbf{q}, \mathbf{q})A_{m\mathbf{k}-\mathbf{q}} )=(\varepsilon_{n\mathbf{k}}-\varepsilon) A_{n\mathbf{k}}.
\end{aligned}
\end{equation}
This expression might be applied to situations when the phase choice breaks time-reversal symmetry or when non-collinear magnetism is present.

\subsection{Variational anisotropic Fr{\"o}hlich model}

In order to apply Eq.~(\ref{eq:23}) to the Fr{\"o}hlich case, this variational framework needs to be reformulated by imposing the model approximations.
At this point one can waive some restrictions of the original model, e.g. consider the case of parabolic energy bands with anisotropic effective masses to get results beyond the classic solutions. The electronic part of the Fr{\"o}hlich Hamiltonian Eq. (\ref{eq:fr_el}) is then
\begin{equation}\label{eq:gfr_el}
    \hat{H}^\text{Fr}_\text{e} = \sum_{\mathbf{k}} \varepsilon({\mathbf{k}}) \hat{c}^\dagger_{\mathbf{k}}\hat{c}_{\mathbf{k}},
\end{equation}
with
\begin{equation}\label{eq:energy}
    \varepsilon({\mathbf{k}}) = \dfrac{1}{2}\left( \dfrac{k^2_x}{m^*_x} + \dfrac{k^2_y}{m^*_y} + \dfrac{k^2_z}{m^*_z} \right).
\end{equation}

We now reformulate Eq.~(\ref{eq:23}) by taking into account the anisotropic Fr{\"o}hlich model approximations.
In the electronic part of Eq.~(\ref{eq:23}) we switch from the KS to the planewave basis, as these are eigenfunctions of the free electron Hamiltonian, so $\psi^0_{n\mathbf{k}}(\mathbf{r})$ in Eq.~(\ref{eq:8}) becomes 
\begin{equation}
    \psi^0_{\mathbf{G}\mathbf{k}}(\mathbf{r}) =  \dfrac{1}{\sqrt{N_p \Omega_0}}e^{i(\mathbf{k}+\mathbf{G})\cdot \mathbf{r}},
\end{equation}
where the band index $n$ now refers to a 3-dimensional index $n_i$ that defines coordinates of a reciprocal lattice vector $\mathbf{G}$. The associated energy given by Eq. (\ref{eq:14}) in this basis reads as
\begin{equation}\label{eq:39}
    E^\text{Fr}_\text{el} = \dfrac{1}{N_p}\sum_{\mathbf{G}\mathbf{k}}|A_{\mathbf{G}\mathbf{k}}|^2\varepsilon({\mathbf{G} + \mathbf{k}}).
\end{equation}

To approach the phonon part we recall that in the Fr{\"o}hlich model only LO phonon mode couples with electrons and we work in the macroscoping limit, ignoring atomic details. The vibrational energy is then modeled by Einstein oscillators with frequency $\omega_\text{LO}$. Taking these approximations into account, in Eq. (\ref{eq:10}) we switch from atoms to Einstein oscillators, each is indexed with $\kappa$ and have mass $M_0$. An oscillator $\kappa$ is moved by a mode $\nu$ only when $\nu = \kappa$, and no coordinate index $\alpha$ is needed, since transverse optical modes are neglected. Thus
\begin{equation}
        \Delta \tau_{\kappa p} = - \dfrac{2}{N_p}\sum_{\mathbf{q}}B^*_{\mathbf{q}\kappa}\left(\dfrac{1}{2M_{0}\omega_\text{LO}}\right)^{1/2}e^{i\mathbf{q}\cdot\mathbf{R}_p}, 
\end{equation}
and one can make an additional Fourier transform to work with the reciprocal lattice vectors $\mathbf{G}$ instead of each index $\kappa$ to characterize the $B^*$ coefficients, since the Einstein oscillators are homogeneously spread:
\begin{equation}\label{eq:41}
    B^*_{\mathbf{q}\kappa} = \frac{1}{\sqrt{N_{\mathbf{G}}}} \sum_{\mathbf{G}} e^{i(\mathbf{q}+\mathbf{G})\cdot\boldsymbol{\tau}_\kappa} B^*_{\mathbf{G}\mathbf{q}},
\end{equation}
where $N_\mathbf{G} = N_\kappa$ is the number of the oscillators, and their homogeneous spread is given by the sum rule
\begin{equation}\label{eq:42}
    \sum_{\kappa}e^{i(\mathbf{G}-\mathbf{G}')\boldsymbol{\tau}_\kappa} = \delta_{\mathbf{G}\mathbf{G}'}N_\kappa.
\end{equation}
From Eqs. (\ref{eq:19}), (\ref{eq:41}) and (\ref{eq:42}) the associated energy is obtained:
\begin{equation}\label{eq:44}
    E_\text{ph}^\text{Fr} = \frac{1}{N_p}\sum_{\mathbf{G}\mathbf{q}}|B_{\mathbf{G}\mathbf{q}}|^2\omega_\text{LO}.
\end{equation}

Lastly, the electron-phonon contribution defined by Eq. (\ref{eq:21}) after some renaming becomes
\begin{equation}\label{eq:45}
\begin{aligned}
    E^\text{Fr}_\text{el-ph} = 
    -\dfrac{1}{N_p^2}\sum_{\mathbf{Gk}}\sum_{\mathbf{G'k'}}A^*_{\mathbf{Gk}}A_{\mathbf{G'k'}}B^*_{(\mathbf{G}-\mathbf{G'}+\mathbf{U})(\mathbf{k}-\mathbf{k'}-\mathbf{U})} \\
    \times g^\text{Fr}(\mathbf{G}-\mathbf{G'} + \mathbf{k}-\mathbf{k'}) + (c.c),
\end{aligned}
\end{equation}
where $\mathbf{U}\equiv\mathbf{U}(\mathbf{k},\mathbf{k}')$ is the Umklapp vector of the reciprocal lattice that translates $\mathbf{q} = \mathbf{k}-\mathbf{k'} - \mathbf{U}$ into the first BZ (possibly $\mathbf{U} = 0$). Electron-phonon matrix elements $g^\text{Fr}(\mathbf{q})$ are given by Eq. (\ref{eq:31}) and (\ref{eq:32}) for a primitive unit cell, since $N_p$ is already present in the prefactor of Eq. (\ref{eq:45}).

In these equations we consider a simple cubic cell with a cubic symmetry and BvK periodic boundary conditions.
One also needs to truncate the summation over $\left( \mathbf{G}, \mathbf{k} \right)$ and $\left( \mathbf{G}, \mathbf{q} \right)$  in the electron and phonon parts respectively.
We note that a uniform $\mathbf{k}$-grid and corresponding $\mathbf{q}$-grid ($\mathbf{q}=\mathbf{k}-\mathbf{k'}-\mathbf{U}$) are simply determined by the size of the BvK supercell.
These grids have to always contain the $\Gamma$-point, but in case when linear size of a grid is even the cubic symmetry will be broken.
On the other hand, electron $\mathbf{G}$-grid and corresponding phonon grid ($\mathbf{G}-\mathbf{G'}+\mathbf{U}$) act as counterparts for definition of electronic bands and phonon modes in Eq. (\ref{eq:23}) and define reciprocal lattice points.
In order to preserve the symmetry and always work with $\Gamma$-centred grids, we utilize the straightforward planewave energy cutoff approach: on the infinite grid for the electron part we select $\left( \mathbf{G}, \mathbf{k} \right)$ with non-zero value of $A_{\mathbf{G} \mathbf{k}}$ only when for a predefined value of $\varepsilon_\text{cut}$
\begin{equation}
    \varepsilon_\text{cut} \ge \varepsilon_{\mathbf{G}\mathbf{k}}.
\end{equation}
Similarly, the phonon coefficients $B_{\mathbf{G}\mathbf{q}}$ are selected only when they connect non-zero electronic coefficients.

Keeping in mind the aforementioned cutoff procedure and combining Eqs. (\ref{eq:39}), (\ref{eq:44}) an (\ref{eq:45}), we arrive at the variational polaron expression applied to the Fr{\"o}hlich model:
\begin{equation}\label{eq:47}
\begin{aligned}
    & \Delta E_\text{p}^\text{Fr}\left( \boldsymbol{A}, \boldsymbol{B} \right) = \dfrac{1}{N_p}\sum_{\mathbf{G}\mathbf{k}}|A_{\mathbf{G}\mathbf{k}}|^2\varepsilon_{\mathbf{G}\mathbf{k}} + \frac{1}{N_p}\sum_{\mathbf{G}\mathbf{q}}|B_{\mathbf{G}\mathbf{q}}|^2\omega_\text{LO} \\
    & -\dfrac{1}{N_p^2}\sum_{\mathbf{Gk}}\sum_{\mathbf{G'k'}}(A^*_{\mathbf{Gk}}A_{\mathbf{G'k'}}B^*_{(\mathbf{G}-\mathbf{G'}+\mathbf{U})(\mathbf{k}-\mathbf{k'}-\mathbf{U})} \\
    &\qquad\qquad\qquad \times g^\text{Fr}(\mathbf{G}-\mathbf{G'} + \mathbf{k}-\mathbf{k'}) + (c.c)).
\end{aligned}
\end{equation}
Minimization of this expression yields polaronic energy in adiabatic approximation, since $\boldsymbol{A} $ and $ \boldsymbol{B} $ are correlated by Eqs. (\ref{eq:24}), (\ref{eq:25}), and strong-coupling scenario of the Fr{\"o}hlich model is captured by this variational approach.
In this case only the quadratic terms of $\Delta E_\text{p}(\alpha)$ expansions given by Eqs. (\ref{eq:35}), (\ref{eq:36}) can be obtained and serve as a benchmark for numerical minimization.

\subsection{Special Treatment of the Fr{\"o}hlich Electron-Phonon Matrix Elements}

Before detailing the minimization algorithm, we note that $g^\text{Fr}(\mathbf{q})$ diverges at $\Gamma$-point. Instead of setting $g^\text{Fr}(0) = 0$, we average this quantity in the neighbourhood of $\Gamma$-point similarly to the approach used in the Supporting Information of Ref. \onlinecite{Miglio2020}.
For this purpose it is convenient to rewrite Eq. (\ref{eq:45})
\begin{equation}\label{eq:rr}
\begin{aligned}
    E^\text{Fr}_\text{el-ph} = 
    -\dfrac{1}{N_p^2}\sum_{\mathbf{Gq}}\left(\sum_{\mathbf{G'k}}A^*_{\mathbf{(G+G'-U)(k+q+U)}}A_{\mathbf{G'k}}\right)B^*_\mathbf{Gq} \\
    \times g^\text{Fr}(\mathbf{G}+\mathbf{q}) + (c.c).
\end{aligned}
\end{equation}
Similar to Eq. (\ref{eq:11}) we obtain the relation between $B_\mathbf{Gq}$ and $A_\mathbf{Gk}$
\begin{equation}
    B_{\mathbf{Gq}}=\dfrac{1}{N_p}\sum_{\mathbf{Gk}}A^*_{(\mathbf{G'+G-U})(\mathbf{k+q+U})}\dfrac{g^\text{Fr}(\mathbf{G'}+\mathbf{q})}{\omega_\text{LO}}A_{\mathbf{Gk}}.
\end{equation}
We also note that exactly at $\mathbf{\Gamma}$-point ($\mathbf{q} = 0, \mathbf{G} = 0$) the term in parenthesis in Eq. (\ref{eq:rr}) equals $N_p$ due to the normalization:
\begin{equation}\label{eq:n}
    \sum_{\mathbf{G'k}}A^*_{\mathbf{G'k}}A_{\mathbf{G'k}} = N_p.
\end{equation}
Now, combining Eqs. (\ref{eq:rr})-(\ref{eq:n}) we express the average contribution of the electron-phonon term of the total energy in the neighborhood of $\Gamma$-point:
\begin{equation}\label{eq:av}
\begin{aligned}
    E_\text{el-ph}^{\mathbf{G}=0, \mathbf{q} \rightarrow 0} \cong - \dfrac{1}{\Omega_{\mathbf{q}=0} N_p^2 \omega_\text{LO}} \int_{\Omega_{\mathbf{q}=0}} \dfrac{d \mathbf{q}}{\mathbf{q}^2}(\mathbf{q} g^\text{Fr}(\mathbf{q}))^2,
\end{aligned}
\end{equation}
where the area of the spherical $\mathbf{q}=0$ neighborhood is denoted by 
\begin{equation}
    \Omega_{\mathbf{q}=0} = \dfrac{4}{3}\pi q_\text{c}^3
\end{equation}
with the cutoff radius
\begin{equation}
    q_\text{c} = 2\pi \left( \dfrac{3}{4\pi\Omega_0} \right)^{1/3}N_p^{-1/3}.
\end{equation}
The term in parenthesis in Eq.~(\ref{eq:av}) is roughly constant and after integration one obtains
\begin{equation}\label{eq:av1}
\begin{aligned}
    E_\text{el-ph}^{\mathbf{G}=0, \mathbf{q} \rightarrow 0} \simeq - \lim_{\mathbf{q}\rightarrow0} \dfrac{(\mathbf{q} g^\text{Fr}(\mathbf{q}))^2}{N_p^2 \omega_\text{LO}} \dfrac{3}{(2\pi)^2}\left( \dfrac{3}{4\pi N_p\Omega_0} \right)^{-2/3}.
\end{aligned}
\end{equation}
Alternatively the same contribution can be obtained if around $\Gamma$-point $g^\text{Fr}(\mathbf{q})$ is replaced by $\overline{g}^\text{Fr}(0)$, which is constant:
\begin{equation}\label{eq:av2}
    \overline{E}_\text{el-ph}^{\mathbf{G}=0, \mathbf{q} \rightarrow 0} \simeq -\dfrac{1}{N_p} \dfrac{\overline{g}^\text{Fr}(0)^2}{\omega_\text{LO}}.
\end{equation}
From Eqs. (\ref{eq:av1}) and (\ref{eq:av2}) one obtains the expression for $\overline{g}^\text{Fr}(0)$. The same procedure can be done in 2D case and corrections to the electron-phonon matrix elements at $\Gamma$-point become
\begin{equation}
\overline{g}_\text{3D}^\text{Fr}(0) =  \dfrac{\sqrt{3}}{2\pi} \left( \dfrac{4\pi N_p \Omega_0}{3} \right)^{1/3} \left( \dfrac{2\pi \omega_\text{LO}}{N_p \Omega_0} {\varepsilon^{*}}^{-1} \right)^{1/2},
\end{equation}
\begin{equation}
\overline{g}_\text{2D}^\text{Fr}(0) = \dfrac{1}{\sqrt{\pi}} \left( \pi N_p \Omega_0 \right)^{1/4} \left( \dfrac{\pi \omega_\text{LO}}{N_p \Omega_0} {\varepsilon^{*}}^{-1} \right)^{1/2}.
\end{equation}
These constants indeed tend to zero for infinitely large supercells, but to a large extent can remove the convergence error of a minimization algorithm at low-density grids. 

\subsection{Preconditioned Conjugate Gradient Algorithm}
\label{ssec:pcg}

The major challenge of the variational approach is the large size of the real space supercell (or equivalently the number of $\mathbf{k}$-points $N_p$) required in the minimization procedure and associated computational complexity.
Since the gradient of the Fr{\"o}hlich variational expression can be easily obtained, see the general Eqs. (\ref{eq:24}) and (\ref{eq:25}), we can utilize an efficient conjugate gradient algorithm and suggest a possible preconditioner to improve the convergence \cite{Pulay1982, Teter1989}.
In Appendix~\ref{asec:pcg} we provide its scaling analysis and show that it has a more favorable scaling than the algorithm utilized by Sio \textit{et al.} in Ref~\citenum{Sio2019}. The present section focuses on the implementation details.

First of all, we note that $\boldsymbol{A}$ and $\boldsymbol{B}$ are linked, as in Eq. (\ref{eq:7}) of the original model:
\begin{equation}\label{eq:53}
\begin{aligned}
    & B_{\mathbf{G''q}}= \\ 
    & \dfrac{1}{N_p}\underbrace{\sum_{\mathbf{Gk}}\sum_{\mathbf{G'k'}}}_{\substack{\mathbf{k-k'-U=q} \\ \mathbf{G-G'+U=G''}}}A^*_{\mathbf{G'}\mathbf{k'}}\dfrac{g^\text{Fr}(\mathbf{G-G'+k-k'})}{\omega_\text{LO}}A_{\mathbf{Gk}}.
\end{aligned}
\end{equation}
This allows one to perform the minimization only in the electronic subspace since the phonon part of the gradient can be always set to zero using Eq (\ref{eq:53}), whatever the value of $\boldsymbol{A}$.
The electronic part of the gradient, which we denote as $\boldsymbol{D}$, is easily obtained from Eq.~(\ref{eq:47}) to give the adaptation of  Eq.~(\ref{eq:24}) to the Fr{\"o}hlich case:
\begin{equation}\label{eq:grad1}
\begin{aligned}
    & D_\mathbf{G'k'} = \dfrac{2}{N_p}A_{\mathbf{G'k'}}(\varepsilon_{\mathbf{G'k'}} - \varepsilon) -\\
    &\dfrac{2}{N_p^2}\sum_{\mathbf{Gk}} (A_{\mathbf{Gk}}B^*_{(\mathbf{G' - G + U})(\mathbf{k' - k - U})}g^\text{Fr}(\mathbf{G'-G+k'-k}) \\
    & + A_{\mathbf{Gk}}B_{(\mathbf{G - G' + U})(\mathbf{k - k' - U})}{g^\text{Fr}}^*(\mathbf{G-G'+k-k'})).
\end{aligned}
\end{equation}

The iterative minimization process itself is as follows.
Let at $n$-th step $\boldsymbol{D}^n$ be the electronic part of the gradient at a certain point.
In order to retain the normalization condition imposed on $\boldsymbol{A}$ we apply the approach similar to Ref.~\citenum{Teter1989}. Firstly, using the Gram-Schmidt process from $\boldsymbol{D}^n$ a vector orthogonal to the $\boldsymbol{A}^n$ is obtained, which we refer to as $\boldsymbol{D}^{\bot n}$.
The conjugate gradient direction is calculated as 
\begin{equation}
    \boldsymbol{Q}^n = \boldsymbol{D}^{\bot n} + \gamma_n \boldsymbol{Q}^{\bot(n-1)}
\end{equation}
with
\begin{equation}
    \gamma_n = \begin{cases}
    0 & n=0 \\
    \dfrac{(\boldsymbol{D}^{\bot n})^* \cdot (\boldsymbol{D}^{\bot n} - \boldsymbol{Q}^{\bot(n-1)})}{|\boldsymbol{Q}^{\bot(n-1)}|^2}  & \text{otherwise}
\end{cases}
\end{equation}
and orthogonalization is also performed to yield $\boldsymbol{Q}^{\bot n}$.

Then the energy is minimized along the path
\begin{equation}
    \Delta E^\text{Fr}(\theta) = \Delta E^\text{Fr}\left(\boldsymbol{A}^n\cos{\theta} + \boldsymbol{Q}^{\bot n}\sin{\theta}\right )
\end{equation}
to find the starting point of the next iteration step
\begin{equation}
\boldsymbol{A}^{(n+1)} = \boldsymbol{A}^n\cos{\theta_\mathrm{min}} + \boldsymbol{Q}^{\bot n}\sin{\theta_\mathrm{min}}.
\end{equation}
The process terminates once the squared norm of the gradient $||\boldsymbol{D}||^2$ becomes lower than a certain threshold. 

A natural preconditioner for the gradient comes from the first term of Eq. (\ref{eq:24}). By taking its inverse for the Fr{\"o}hlich case one obtains the preconditioner
\begin{equation}
    {P}_{\mathbf{G}\mathbf{k}} = N_p(\varepsilon_{\mathbf{G}\mathbf{k}} - \varepsilon_\mathrm{mod})^{-1},
\end{equation}
where $\varepsilon_\mathrm{mod}$ is the fraction of $\varepsilon$.
While the optimal choice of $\varepsilon$ can be a challenge in full first-principles calculations, since the target value of this quantity cannot be estimated before the minimization, Pekar's 1:2:3:4 theorem allows its precise definition for the standard Fr{\"o}hlich model and gives qualitative estimation in case of generalized Fr{\"o}hlich model. 
Hence instead of simple gradient $\boldsymbol{D}$, one can use the preconditioned gradient $\boldsymbol{D}^{\text{PC}}$, with components  ${D}^{\text{PC}}_{\mathbf{G}\mathbf{k}}={P}_{\mathbf{G}\mathbf{k}}{D}_{\mathbf{G}\mathbf{k}}$, to reach the solution significantly faster. 

We also note that if Eq. (\ref{eq:47}) is minimized at $\boldsymbol{A}$, it reaches the minimum at $\boldsymbol{A}^*$ as well, which implies that electronic coefficients are real-valued. 
In addition the polaron wavefunction inherits the symmetries of the problem and this allows a reduction of computational time and memory by evaluating $\boldsymbol{A}$, $\boldsymbol{B}$ and $g^\text{Fr}(\mathbf{q})$ only in the irreducible part of the BZ defined by its symmetries.

Finally, since the minimum of $\Delta E_p$ is obtained for a finite supercell defined by the size of a $\mathbf{k}$-grid, we make a series of optimizations for several grids and then perform an extrapolation similar to the Makov-Payne extrapolation \cite{Makov1995} to obtain the polaron formation energy in the infinite limit $\Delta E_p^\infty$:
\begin{equation}
    \Delta E_\text{p}(N_p) = \Delta E_\text{p}^\infty + a N_p^{-1} + \mathcal{O}(N_p^{-n}),
\end{equation}
where $N$ is the linear size of a grid and $n=2/3$ in 2D/3D~case. The leading size-dependent term is coherent with Eq.~\ref{eq:av2}.

\section{Results and Discussion}
\label{result}

\begin{figure}[t]
\includegraphics[scale=0.35]{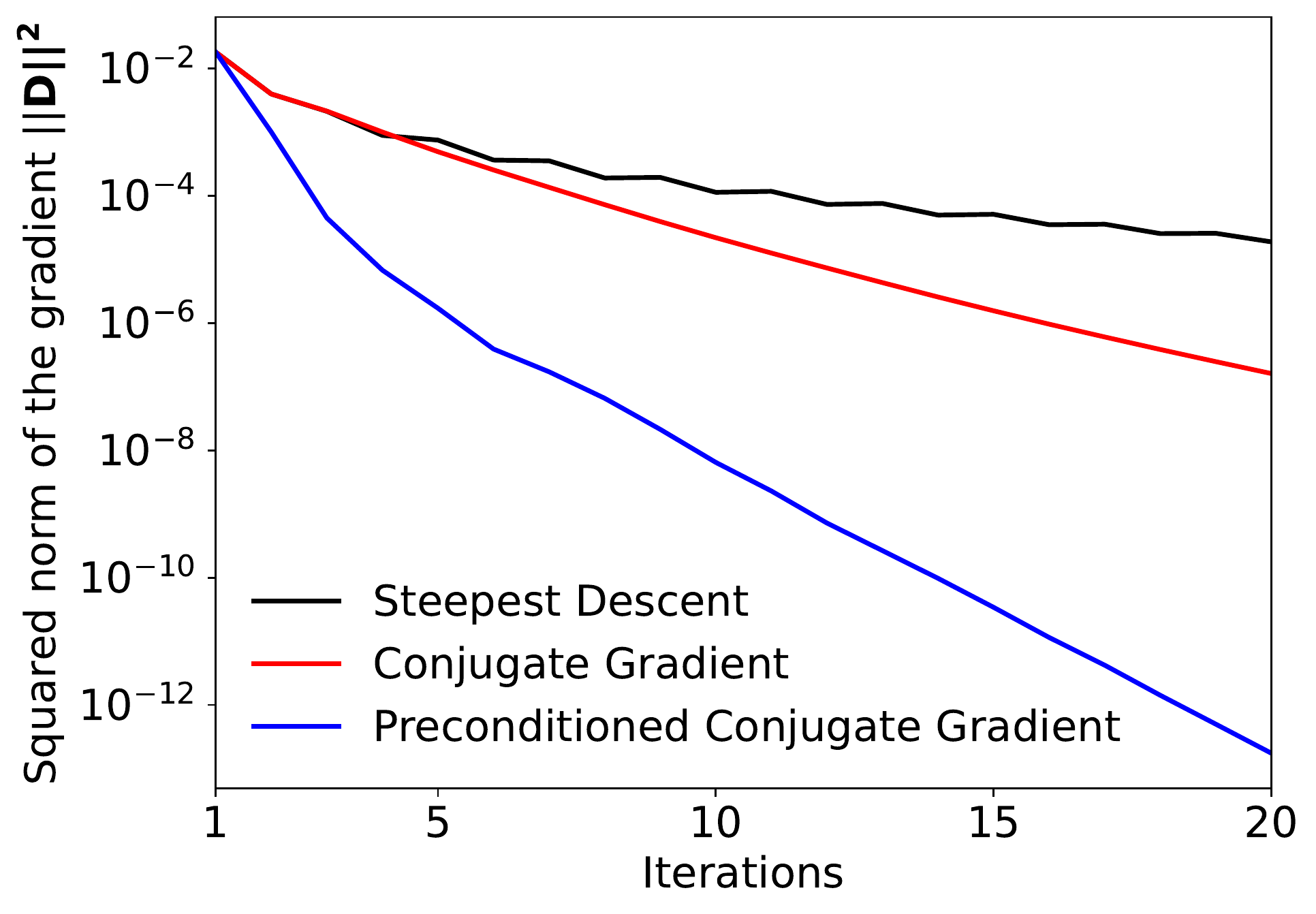}
\caption{Performance comparison of gradient-based optimization algorithms applied to the variational Fr{\"o}hlich model. The model is 2D isotropic with $m^*=1$, $\varepsilon^*=1$ and $20\times20~\mathbf{k}\text{-point grid}$.}
\label{fig:alg_comparison}
\end{figure}

We begin the analysis of the variational Fr{\"o}hlich model by comparing the efficiency of different gradient-based algorithms applied to the polaron energy minimization. Along with the preconditioned conjugate gradient (PCG) mentioned in Section \ref{ssec:pcg} we also consider conjugate gradient wihout preconditioning (CG) and steepest descent (SD), which are obtained from PCG by setting ${P}_{\mathbf{G}\mathbf{k}}=1$ and also $\gamma_n$ = 0 in case of SD.
Fig. \ref{fig:alg_comparison} shows that PCG decreases the squared norm of the gradient $||\boldsymbol{D}||^2$ most rapidly, and the other two methods are substantially slower. This behaviour is consistent for various range of the model parameters and thus PCG is the best choice for optimization. 
It should be noted that in the original SVPG paper\cite{Sio2019a} authors employ a parallel SD method, but even in such general first-principles model, implementation of PCG can be a major improvement that increases convergence rates of the iterative minimization, see Appendix \ref{asec:pcg}. 

\begin{figure}[t]
\includegraphics[scale=0.315]{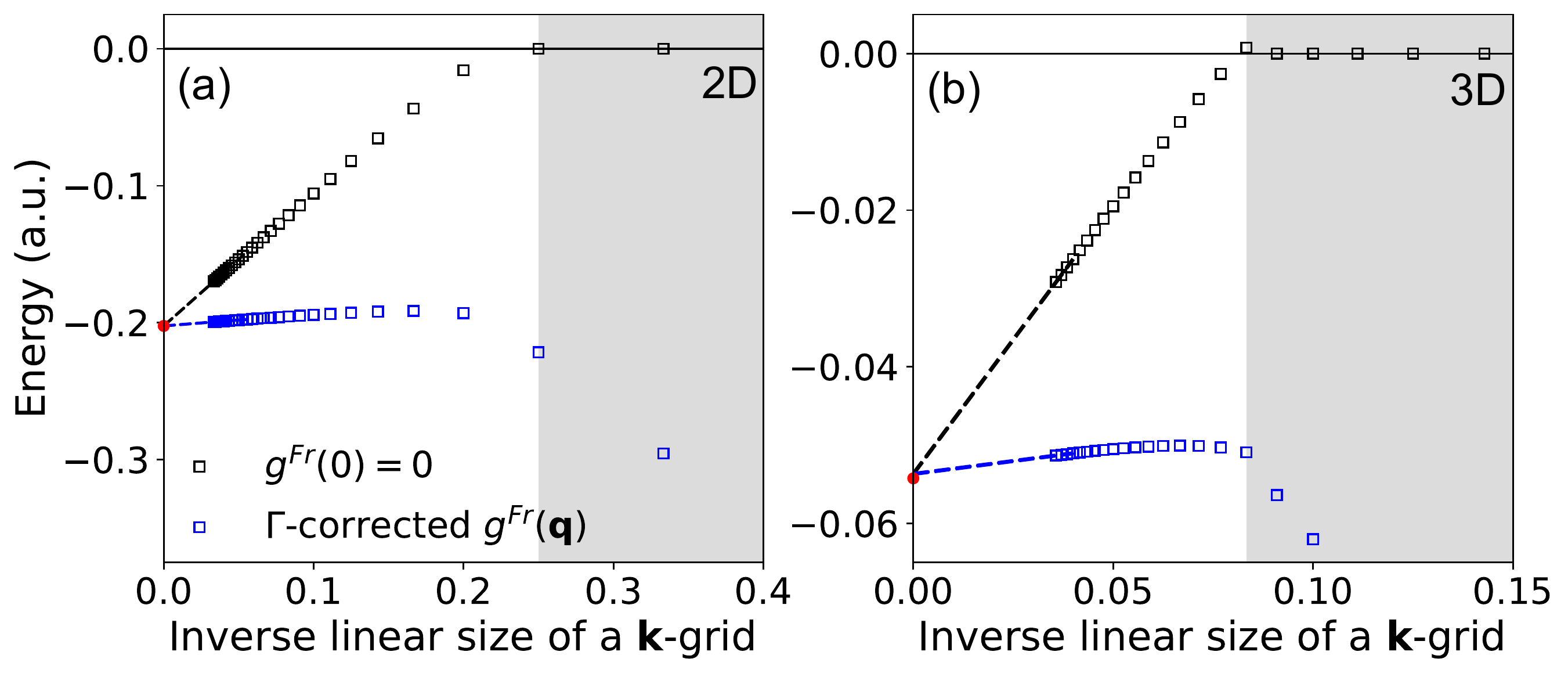}
\caption{Polaron formation energy $\Delta E_p$  for the (a) 2D and (b) 3D isotropic Fr{\"o}hlich model with $m^*=1$, $\epsilon^*=1$, 
as a function of the inverse linear size of the wavevector sampling. The square markers denote $\Delta E_\text{p}$ obtained after minimization for range of $\mathbf{k}$-point grids of incrementally increasing density up to $30\times30$ and $28\times28\times28$ in 2D and 3D cases respectively.  Black points are obtained by setting $g^\text{Fr}(0) = 0$, while Eq.~\ref{eq:av2} is used for the blue points. The dashed lines are the extrapolation and the red circles show asymptotic solutions of the original Fr{\"o}hlich model. The grey regions indicate ranges of $\mathbf{k}$-point grids for which no polaron is formed.}
\label{fig:alg_makovpayne}
\end{figure}

In order to validate the optimization results we compare polaron formation energies obtained in the isotropic case with the asymptotic solution of the Fr{\"o}hlich model. 
Fig. \ref{fig:alg_makovpayne} shows that the extrapolation yields $\Delta E^\infty_\text{p}$ that is in agreement with the asymptotic solution with the small difference due to the finite size of the $\mathbf{k}$-grids employed.
It is the leading term $\gamma = \Delta E^\infty_\text{p}/(\alpha^2\omega_\text{LO})$ (strong-coupling coefficient) in Eqs. (\ref{eq:35}) and (\ref{eq:36}) that can be obtained in the adiabatic strong-coupling approximation,
and calculations give $\gamma  = -0.4046$ in 2D and $\gamma = -0.1074$ in 3D, while the reference values are $-0.4047$ and $-0.1085$ respectively.
Corrections of $g^\text{Fr}(\mathbf{q})$ at $\Gamma$ do not affect the extrapolated energy value since they vanish when $N_p \rightarrow \infty$. 
However, at low-density $\mathbf{k}$-point grids they allow to obtain formation energies much more accurate than the calculations done with $g^\text{Fr}({0}) = 0.$ 
Thus one can qualitatively estimate the value of $\Delta E^\infty_\text{p}$ already with a small $\mathbf{k}$-grid without an extrapolation.
Additionaly, there exist $\mathbf{k}$-point grids or alternatively supercells of critical size up to which there is no polaron formation. 
This behavior is similar to the one that can be found in Ref. \citenum{Sio2019a, Sio2019} and is explained by the transition from a delocalized electronic state to a localized (self-trapped) state, i.e. periodic images of the polaron interact and form an extended wavefunction if a supercell is too small, so the quasiparticle is fully delocalized.
\begin{figure}[h]
\includegraphics[scale=0.31]{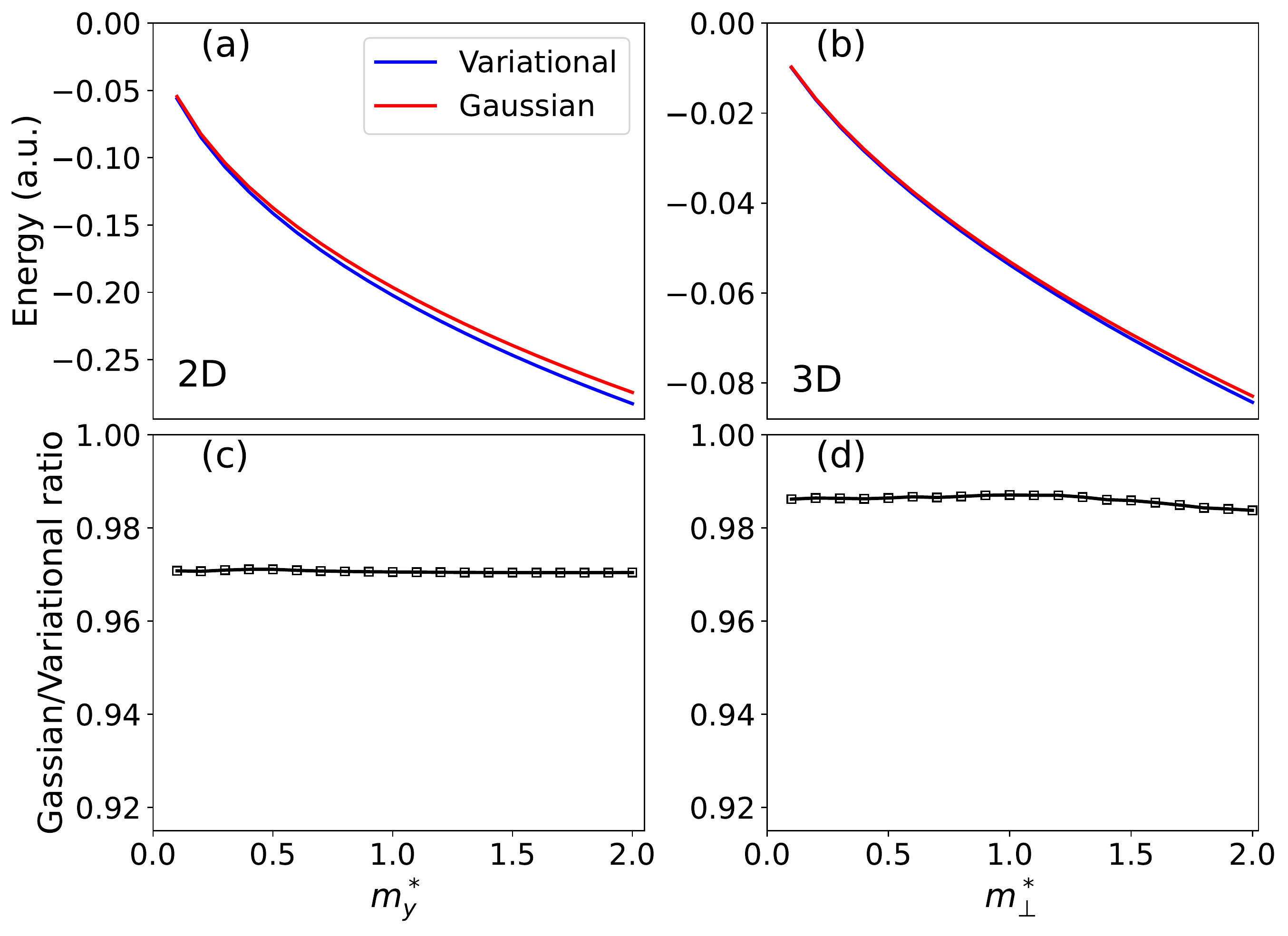}
\caption{Comparison between the (a), (c) 2D and (b), (d) 3D generalized variational Fr{\"o}hlich model and Gaussian ansatz approach in case of anisotropic electronic bands. Effective mass is fixed along a preferred direction ($m^*_x = 1$) and varied along the perpendicular ones (in 3D $m^*_\bot \equiv m^*_y = m^*_z$). Top panels represent the dependence of the extrapolated polaron formation energies $\Delta E^\infty_\text{p}$ on the effective masses, and bottom panels show the ratio between the Gaussian  and variational results, taking the latter as reference.}
\label{fig:vargau}
\end{figure}
\begin{figure}[h]
\includegraphics[scale=0.335]{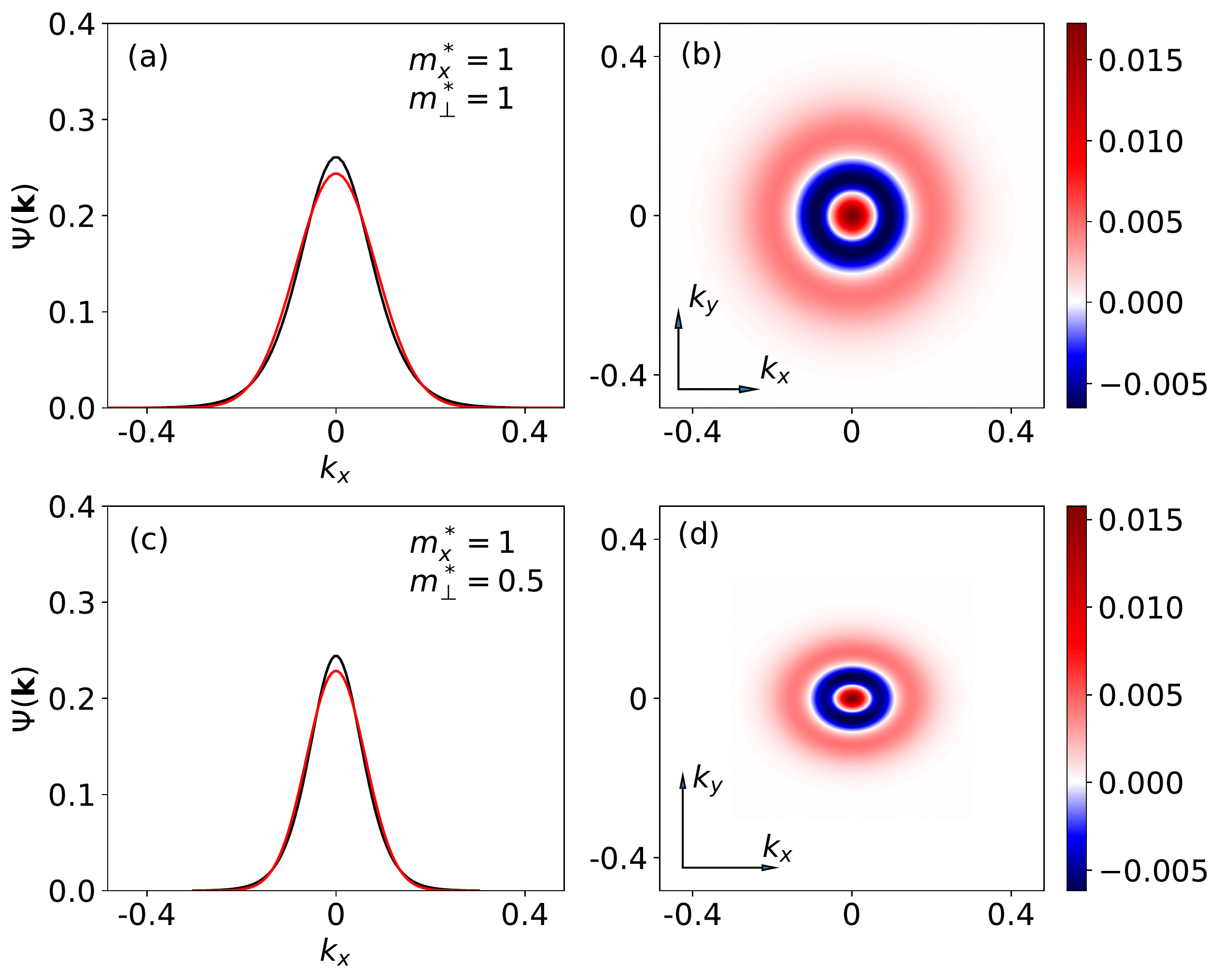}
\caption{Cross-sections (a) and (c) of the numerically exact (black line) and optimized Gaussian (red line) wavefunctions along the $k_x$~direction in reciprocal space with different anisotropy; (b) and (d) indicate the $k_x k_y$-plane cross-sections of the wavefunctions difference.}
\label{fig:wavefunction}
\end{figure}

Next we examine the generalized 2D and 3D Fr{\"o}hlich model when electronic bands are anisotropic and all the other approximations of the original Fr{\"o}hlich model remain valid.
We also compare our results with the ones obtained with the Gaussian ansatz approach used by Guster \textit{et al}~\cite{Guster2021}, which treats polarons in the strong-coupling adiabatic approximation like in the current methodology.
As shown in Fig.~\ref{fig:vargau}, the difference is rather small, and quite independent of the anisotropy in the shown anisotropy parameter range, with the variational approach giving only up to 3 \% more accurate results.
Taking into account numerical errors, the ratio between the two methods is constant in 2D and 3D cases, which implies that numerically exact polaronic wavefunction deviates from the optimized Gaussian trial wavefunction in a consistent manner regardless of the effective mass ratio as shown in Fig.~\ref{fig:wavefunction}.
However, the deviation may become more pronounced once further generalizations to the model are introduced (degeneracy, multiple phonon bands), but this requires additional investigation and is not in the scope of the current work.

Finally, we look at the individual terms that contribute to the polaron formation energy ($E_\text{el}$, $E_\text{ph}$, $E_\text{el-ph}$) as well as the eigenenergy of its localized state $\varepsilon$.
We observe that in both anisotropic and isotropic 2D and 3D cases Pekar's 1:2:3:4 theorem remains valid with only small deviations due to numerical inaccuracies. 
In Tables \ref{tab:2d_1234} and \ref{tab:3d_1234}, we report these values, divided by the corresponding Pekar coefficient:  $E_\text{el}$, $E_\text{ph}/2$, $\varepsilon/3$ and $E_\text{el-ph}/4$. We call these quantities the ``reduced" energies.
This behavior also supports accuracy of the obtained results and can be used as a convergence check when the extrapolation is performed: $\mathbf{k}$-point grid density and energy cutoff $\varepsilon_\text{ecut}$ are good enough if the Pekar's 1:2:3:4 theorem is fulfilled, and have to be increased otherwise.
Additional discussion on the validity of this theorem in anisotropic case is also provided in Appendix \ref{asec:pekar}.
\begin{table}[t]
\centering
\renewcommand\arraystretch{1.5}
\caption{2D generalized Fr{\"o}hlich model. Absolute values of reduced energies : polaron formation energy $\Delta E_\text{p}$, its decomposition into individual (reduced) electron $E_\text{el}$, phonon $E_\text{ph}/2$, electron-phonon $E_\text{el-ph}/4$ terms, and localized polaronic state reduced eigenenergy $\varepsilon/4$, ordered as in Pekar 1:2:3:4 theorem.}
\label{tab:2d_1234}
\begin{tabular}{cccccc}\hline \hline
$m^*_y$ & $\Delta E_\text{p}$ & $E_\text{el}$ & $E_\text{ph}/2$ & $\varepsilon/3$ & $E_\text{el-ph}/4$  \\ \hline
1.0 &	0.2023 &	0.2026 &	0.2024 &	0.2024 &	0.2024 \\
0.8 &	0.1806 &	0.1811 &	0.1809 &	0.1808 &	0.1809 \\
0.6 &	0.1556 &	0.1561 &	0.1558 &	0.1557 &	0.1558 \\
0.4 &	0.1250 &	0.1220 &	0.1235 &	0.1240 &	0.1235 \\
0.2 &	0.0848 &	0.0849 &	0.0848 &	0.0848 &	0.0848 \\ 
\hline \hline
\end{tabular}
\end{table}
\begin{table}[t]
\centering
\renewcommand\arraystretch{1.5}
\caption{3D generalized Fr{\"o}hlich model. Absolute values of reduced energies : polaron formation energy $\Delta E_\text{p}$, its decomposition into individual (reduced) electron $E_\text{el}$, phonon $E_\text{ph}/2$, electron-phonon $E_\text{el-ph}/4$ terms, and localized polaronic state reduced eigenenergy $\varepsilon/4$, ordered as in Pekar 1:2:3:4 theorem.}
\label{tab:3d_1234}
\begin{tabular}{cccccc}\hline \hline
$m^*_\bot$ & $\Delta E_\text{p}$ & $E_\text{el}$ & $E_\text{ph}/2$ & $\varepsilon/3$ & $E_\text{el-ph}/4$  \\ \hline
1.0 &	0.0537 &	0.0549 &	0.0543 &	0.0541 &	0.0543 \\
0.8 &	0.0463 &	0.0465 &	0.0464 &	0.0463 &	0.0464 \\
0.6 &	0.0379 &	0.0413 &	0.0396 &	0.0390 &	0.0396 \\
0.4 &	0.0284 &	0.0282 &	0.0283 &	0.0283 &	0.0283 \\
0.2 &	0.0168 &	0.0163 &	0.0166 &	0.0167 &	0.0166 \\
\hline \hline
\end{tabular}
\end{table}


\section{Conclusion}
\label{conclusion}
In this work, starting from recent advances in the first-principles modeling of polarons by Sio \textit{et al}\cite{Sio2019a, Sio2019} we derive variational polaron equations in the basis of Kohn-Sham states.
We suggest an effective gradient-based optimization algorithm and apply it to the Fr{\"o}hlich model in 2D and 3D. 
We compare obtained results with the known isotropic asymptotic solution, and observe an excellent agreement. 
We also investigate the case of the anisotropic Fr{\"o}hlich model, showing that the full variational approach gives slightly more accurate solution than the Gaussian ansatz technique.
Apart from that, the divergent Fr{\"o}hlich electron-phonon matrix elements are corrected at the $\Gamma$-point, reducing by a large factor the convergence error, and allowing for qualitative estimation of the polaron formation energy without any extrapolation.
Our methodology also allows to obtain the energy of a localized polaronic state and decompose the formation energy into individual electronic, vibrational and electron-phonon contributions.
We show that their ratio obey's Pekar's 1:2:3:4 rule regardless of anisotropy and dimensionality. 

While the main application of the current work is on the anisotropic Fr{\"o}hlich model, further generalization can be performed using the suggested framework.
Taking also into account possible degeneracy of electronic bands as well as several LO phonon modes one may study wide range of realistic materials in scope of the generalized Fr{\"o}hlich model.

\begin{acknowledgments}
This work has been supported by the Fonds de la Recherche Scientifique (FRS-FNRS, Belgium) through PdR ALPS Grant No. T.0103.19 .
\end{acknowledgments}

\appendix
\renewcommand\thefigure{A\arabic{figure}}
\section{Electron-phonon part of the Hamiltonian with arbitrary choice of phonon phase}\label{asec:phase}
The general phase relation for the phonon eigenmodes with opposite wavevectors, obtained from the diagonalization of dynamical matrix (e.g. by numerical means) reads as
\begin{equation}
     e_{\kappa\alpha, \nu}(-\mathbf{q}) = e^{i\phi(\mathbf{q})} e^*_{\kappa\alpha, \nu}(\mathbf{q}),
\end{equation}
with $\phi(\mathbf{q})$ being an arbitrary phase. This equation is valid for non-degenerate phonon states, and should be further generalized to unitary matrices for the degenerate case, although we will not treat this further generalization in the present appendix. Actually, the phase $e^{i\phi(\mathbf{q})}$ depends on the mode $\nu$, but for the sake of simplicity, we will not explicitly mention this dependence.

There are two convenient conventions for the choice of phase, namely $\phi(\mathbf{q}) = 0$ as in Born and Huang \cite{Born54} and $\phi(\mathbf{q}) = \pi$ as in Leibfried. \cite{Leibfried55} The first, for example, is used in Ref. \citenum{Giustino2017}. However, without choosing any of these conventions, one can obtain a generalized expression for a linear coordinate transformation of the ionic displacements and the accompanying electron-phonon term in the Hamiltonian. 

For this purpose, we follow Appendix B of Ref. \citenum{Giustino2017}, which delivers the following modified equations. Eq. (B15) of Ref. \citenum{Giustino2017} becomes
\begin{equation}\label{eq:z}
    z_{\mathbf{q}\nu} = l_{\mathbf{q}\nu}(\hat{a}_{\mathbf{q}\nu} + e^{-i\phi(\mathbf{q})}\hat{a}^\dagger_{-\mathbf{q}\nu}).
\end{equation}
Using this equation one gets atomic displacements
\begin{equation}
\begin{aligned}
    & \Delta \tau_{\kappa\alpha p} = \\
    & \left( \dfrac{M_0}{N_p M_\kappa} \right)^{1/2} \sum_{\mathbf{q} \nu} e^{i\mathbf{q}\cdot\mathbf{R}_p}  e_{\kappa\alpha, \nu}(\mathbf{q}) l_{\mathbf{q}\nu}(\hat{a}_{\mathbf{q}\nu} + e^{-i\phi(\mathbf{q})}\hat{a}^\dagger_{-\mathbf{q}\nu})
\end{aligned}
\end{equation}
and electron-phonon term of the Hamiltonian
\begin{equation}\label{eq:hfr}
\begin{aligned}
    & \hat{H}_\text{el-ph} = \\
    & \dfrac{1}{N_p^{1/2}} \sum_{\substack{\mathbf{kq} \\ mn\nu}}g_{mn\nu}(\mathbf{k}, \mathbf{q})\hat{c}^\dagger_{m\mathbf{k}+\mathbf{q}}\hat{c}_{n\mathbf{k}} (\hat{a}_{\mathbf{q}\nu} + e^{-i\phi(\mathbf{q})}\hat{a}^\dagger_{-\mathbf{q}\nu}),
\end{aligned}
\end{equation}
which are the sought generalized counterparts of Eqs. (20) and (37) of Ref. \citenum{Giustino2017} respectively.

Eq. (\ref{eq:hfr}) shows that the form of the  Fr{\"o}hlich Hamiltonian depends on the choice of phase. In the present paper $\phi(\mathbf{q}) = 0$, but if Leibfrid convention were used, Eqs.~(\ref{eq:fr_elph})-(\ref{eq:32}) would read as
\begin{equation}
    \hat{H}^\text{Fr}_\text{el-ph} = \sum_{\mathbf{k}\mathbf{q}} g^\text{Fr}(\mathbf{q})\hat{c}^\dagger_{\mathbf{k}+\mathbf{q}}\hat{c}_{\mathbf{k}}(\hat{a}_{\mathbf{q}} - \hat{a}^\dagger_{\mathbf{-q}}),
\end{equation}
\begin{equation}
g_\text{3D}^\text{Fr}(\mathbf{q}) = \dfrac{i}{|\mathbf{q}|}\left( \dfrac{2\pi \omega_\text{LO}}{N_p \Omega_0} {\epsilon^{*}}^{-1} \right)^{1/2},
\end{equation}
\begin{equation}
g_\text{2D}^\text{Fr}(\mathbf{q}) = \dfrac{i}{|\mathbf{q}|^{1/2}}\left( \dfrac{\pi \omega_\text{LO}}{N_p \Omega_0} {\epsilon^{*}}^{-1} \right)^{1/2},
\end{equation}
which is coherent, e.g. with Refs. \citenum{Verdi2015, devreese2000polarons}. See Ref.~\onlinecite{Guster2022} for further information about this topic.

\section{Pekar's 1:2:3:4 theorem}\label{asec:pekar}

In order to derive the 1:4 relation of the Pekar's theorem in anisotropic case it is convenient to introduce the average effective mass $\overline{m^*}=\sqrt[3]{m^*_x m^*_y m^*_z}$, and rewrite the electronic energy given by Eq. (\ref{eq:energy}) as follows :
\begin{equation}\label{eq:energy1}
    \varepsilon({\mathbf{k}}) = \dfrac{1}{2\overline{m^*}}\left( \dfrac{k^2_x}{m^*_{x,r}} + \dfrac{k^2_y}{m^*_{y,r}} + \dfrac{k^2_z}{m^*_{z,r}} \right),
\end{equation}
where $m^*_{i,r} =
m^*_{i} / \overline{m^*} $ denotes reduced effective masses along each direction. These reduced effective masses will stay unchanged in what follows, which is a key point in the demonstration.

Then one can follow the same reasoning provided in Ref.~\citenum{Lemmens1973} for isotropic case.
Assuming $\psi^0$ is the ground-state wavefunction of a polaron in the ground state, we use Fr{\"o}hlich Hamiltonian defined by Eq. (\ref{eq:27}) and Feynman-Hellman theorem to obtain the derivative of the polaron formation energy with respect to inverse of the mass $\lambda = 1/\overline{m^*}$:
\begin{equation}\label{eq:der1}
    \dfrac{d (\Delta E_\text{p})}{d \lambda} = \left< \psi^0 \left| \dfrac{d H^\text{Fr}}{d \lambda} \right| \psi^0 \right> = \dfrac{1}{\lambda} E_\text{el}.
\end{equation}
After changing the Hamiltonian to dimensionless units similarly to Ref.~\citenum{Lemmens1973}, it can be shown that the polaron formation energy $\Delta E_\text{p}$ is the function of $\overline{\alpha}$ only, which is the averaged anisotropic Fr{\"o}hlich coupling constant:
\begin{equation}
    \overline{\alpha} = \left( \dfrac{\overline{m^*}}{2\omega_\text{LO}} \right)^{1/2}{\epsilon^{*}}^{-1}.
\end{equation}
Then using the average mass dependence of this constant and Eq.~(\ref{eq:der1}), electronic contribution to the formation energy can be expressed as derivative of $\overline{\alpha}$:
\begin{equation}\label{eq:der2}
    E_\text{el} = \lambda \dfrac{d (\Delta E_\text{p})}{d \lambda} = \lambda \dfrac{d (\Delta E_\text{p})}{d \overline{\alpha}} \dfrac{d \overline{\alpha}}{d \lambda} = 
    -\dfrac{1}{2} \overline{\alpha} \dfrac{d (\Delta E_\text{p})}{d \overline{\alpha}}.
\end{equation}
The electron-phonon interaction term, obtained in the same way, reads as
\begin{equation}\label{eq:der3}
    E_\text{el-ph} = 2\overline{\alpha} \dfrac{d (\Delta E_\text{p})}{d \overline{\alpha}}.
\end{equation}
Combining Eqs. (\ref{eq:der2}) and (\ref{eq:der3}) one obtains the 1:4 relation of the Pekar's theorem, that is valid regardless of the value of $\overline{\alpha}$:
\begin{equation}\label{eq:der4}
    E_\text{el} : -E_\text{el-ph} = 1:4.
\end{equation}

Since the present work describes polaron in the strong coupling regime of adiabatic approximation, the 1:2 relation will always hold. This immediately follows from the fact that electron and phonon parts of the polaron wavefunctions are correlated, see Eq. (\ref{eq:25}). Taking this into account and using Eqs. (\ref{eq:19}), (\ref{eq:22}), one can show that
\begin{equation}
    E_\text{el-ph} = -2E_\text{ph}
\end{equation}
and, consequently, from the relation \ref{eq:der4} it follows that
\begin{equation}
    E_\text{el} : E_\text{ph} = 1:2.
\end{equation}

Lastly, adiabatic approximation implies that the formation energy of a self-trapped polaron can also be expressed as the sum of the energy of localized polaronic state and the energy of lattice deformation:
\begin{equation}\label{eq:aen}
    \Delta E_\text{p} = \varepsilon + E_\text{ph}.
\end{equation}
Using Eq.~(\ref{eq:13}) and the aforementioned ratios, one obtains the 1:3 relation:
\begin{equation}
    E_\text{el} : -\varepsilon = 1:3.
\end{equation}

It should be emphasized that the 1:2 and 1:3 ratios are inherent to the adiabatic approximation and will always be valid for the model of Ref.~\citenum{Sio2019} irrespective of the form of the electron-phonon coupling.
On the other hand, the 1:4 ratio comes from the dimensional analysis of the Fr{\"o}hlich model, which arises when one takes derivatives of the formation energy with respect to a certain parameter.
Hence in the present work the 1:4 result is only true when the convergence with respect to number of plane waves and number of $\mathbf{k}$-points is reached.

In the weak-coupling regime, only the 1:4 relation remains.
Vibrational energy $E_\text{ph}$ in this case will depend on the $\overline{\alpha}$ parameter, and energy of a localized polaronic state $\varepsilon$ is not even defined, since there is no self-trapped polaron.
In this sense we note that in Appendix B of Ref.~\citenum{Kokott2018} the authors erroneously state that the Pekar's theorem is valid for all ranges of the coupling parameter.

\section{Scaling of the Minimization Algorithm}\label{asec:pcg}
\renewcommand\thefigure{\thesection.\arabic{figure}}
\setcounter{figure}{0} 
\begin{figure}[h]
\includegraphics[scale=0.30]{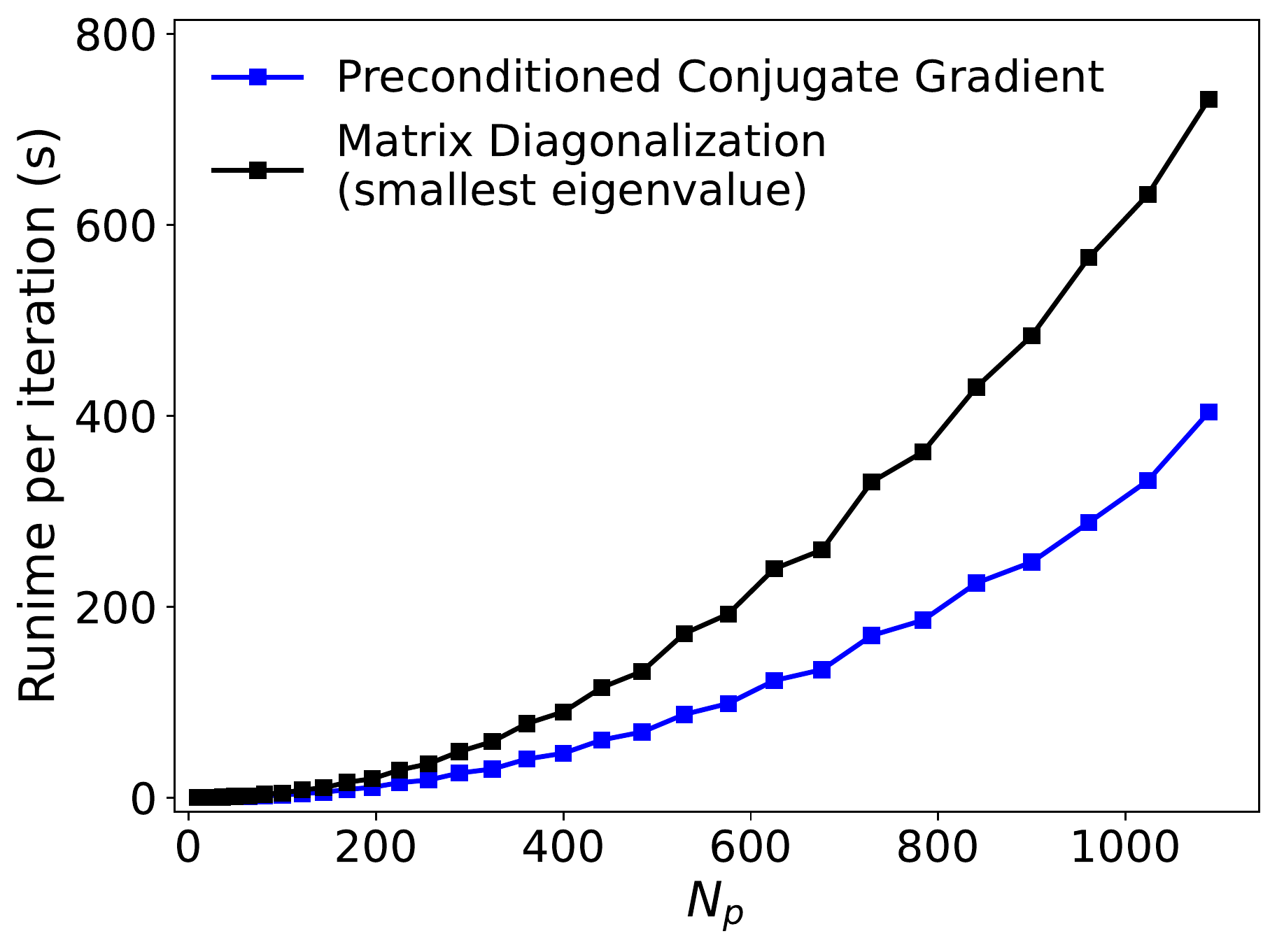}
\caption{Comparison between the runtime of minimization step for an iterative eigensolver and conjugate gradient descent algorithm.}
\label{fig:complexity}
\end{figure}

The computational complexity of the SVPG framework and, consequently, of the variational approach to the problem is determined by the large size of a $\mathbf{k}$-point grid $N_p$ that may be required for the extrapolaton of energy.
In the original papers authors rely on standard numerical eigensolvers to diagonalize electronic matrix defined by Eq.~(\ref{eq:6}), which scale like $\mathcal{O}(N_p^3)$, i.e. as the cube of the matrix size.
In their approach only the lowest (largest) eigenvalue of the Hamiltonian is required, so to find this value and the corresponding eigenvector one can also benefit from iterative eigensolvers that scale like $\mathcal{O}(N_p^2)$.

Computation of the gradient at each minimization step defined by Eqs.~(\ref{eq:24}),~(\ref{eq:25}) scales as $\mathcal{O}(N_p^2)$, which is also valid in the Fr{\"o}hlich case, see Eqs.~(\ref{eq:53}) and (\ref{eq:grad1}).
This along with the efficiency of the PCG algorithm allowed us to handle the calculations on a laptop using Python \cite{10.5555/1593511} scripts, while in Ref.~\citenum{Sio2019} authors rely on a distributed-memory eigensolver from the ScaLAPACK library \cite{CHOI19961} to deal with grids of similar size.
Also, in the matrix methods setting up the Hamiltonian requires additional memory in comparison with the variational approach.
Fig.~\ref{fig:complexity} shows performance of the minimization~routines.

\newpage
\bibliography{main} 

\end{document}